# Gradient index phononic crystals and metamaterials

Yabin Jin[1,*], Bahram Djafari-Rouhani[2,*], Daniel Torrent[3,*]

[1] School of Aerospace Engineering and Applied Mechanics, Tongji University, 200092, Shanghai, China

[2] Institut d'Electronique, de Microélectonique et de Nanotechnologie, UMR CNRS 8520, Département de Physique, Université de Lille, 59650 Villeneuve d'Ascq, France

[3] GROC-UJI, Institut de Noves Tecnologies de la Imatge, Universitat Jaume I, 12071, Castello, Spain

[*]Corresponding authors:

083623jinyabin@tongji.edu.cn;

bahram.djafari-rouhani@univ-lille.fr;

dtorrent@uji.es

**Abstract**

Phononic crystals and acoustic metamaterials are periodic structures whose effective properties can be tailored at will to achieve extreme control on wave propagation. Their refractive index is obtained from the homogenization of the infinite periodic system, but it is possible to locally change the properties of a finite crystal in such a way that it results in an effective gradient of the refractive index (GRIN). In such case the propagation of waves can be accurately described by means of ray theory, and different refractive devices can be designed in the framework of wave propagation in inhomogeneous media. In this paper we review the different devices that have been studied for the control of both bulk or guided acoustic waves based on graded phononic crystals.

## 1. Introduction

Phononic crystals and acoustic metamaterials enable to achieve innovative properties for the propagation of mechanical waves (air-borne sound waves, water-borne acoustic waves, water waves, elastic waves, surface acoustic waves, Lamb waves) inapproachable in natural materials. Phononic crystals consist of periodic arrangements of scatterers in a given matrix, and they were firstly proposed in 1993[1, 2]. The attention that phononic crystals received originally was due to the existence of a phononic Bragg band gap where the propagation of acoustic waves was forbidden.

In this regime the wavelength λ of the acoustic field is comparable to the periodicity *a* of the lattice, λ≈*a*, and to be observable it is also required that the thickness *D* of the bulk phononic crystals be at least four or five periodicities, *D*>>*a*. In the subwavelength range, λ>>*a* , it is possible to find local resonances in the scatterers and the designed structures exhibit hybridization band gaps which give rise to novel effects such as negative mass density or negative elastic modulus. In this exotic regime the structure behaves as a special type of materials called "acoustic metamaterials", which were firstly proposed in the seminal work[3] in 2000. Over the past two decades, dramatically increasing efforts have been devoted to the study of acoustic artificial structured materials driven by both fundamental scientific curiosities with properties not found previously and diverse potential applications with novel functionalities[4-16].

While interesting, most of the extraordinary properties of metamaterials are in general single-frequency or narrow-band, since outside the resonant regime metamaterials behave as common composites. However, non-resonant phononic crystals in the low frequency regime behave as homogeneous non-dispersive materials whose effective parameters can be easily tailored, and in this regime gradient index (GRIN) acoustic materials, or GRIN devices, are easily doable. These devices were firstly proposed for acoustic waves by Torrent et al in 2007[17] and for elastic waves by Lin et al in 2009[18], and they allow to manipulate acoustic waves to enforce them to follow curved trajectories. They are characterized by a spatial variation of acoustic refractive index, which is designed by locally changing the geometry of units. For instance, the effective index obviously depends on the filling ratio of scatterers, namely the size of scatterers, which is initially proposed for gradient index control of acoustic waves[17-19]; instead, the variation in the lattice spacing while keeping the size of scatterers is also achievable[20, 21]; for triangular shape of scatterers, rotating the angles of the triangular shape can also affect the effective acoustic velocity[22]; for lead-rubber pillared metamaterial plate, by changing the height of lead layer in pillars, the effective mass density is tuned, resulting in a change in effective phase velocity following a given law [23]; for phononic crystal plates, the effective velocity of antisymmetric Lamb wave is directly related with the thickness of plates, which can also further affect the effective velocity of symmetric mode [24-26]. In addition to the geometric parameters, the effective refractive index can also be tuned in relation to the elastic properties of the scatterers, for instance  the choice of materials[18, 27], or to an external stimulus such as electric field[28], or temperature[29].

The attained effective refractive index of the GRIN device can be smaller or larger than that of the background medium, which corresponds to phase advance[20] or delay approaches in wave propagation. Nevertheless, the phase delay approach is mostly employed as higher refractive index enables to design more advanced functionalities and reduce the entire width of the devices. Comparing to the wavelength, the thinner is the device, the larger will be the highest required index. If the width of the whole device is downscaled to the subwavelength regime, the GRIN device becomes a metasurface[30, 31]. However, to find a high enough effective refractive index to design GRIN metasurfaces is a big challenge that needs new technologies in material science. Recently, it has been reported that soft porous materials[32, 33] with soft-matter techniques can achieve a relative refractive index higher than 20[34]. GRIN devices can also be designed with negative index of refraction at frequencies laying in the first negative slope of the acoustic band structure[21]. However, these devices would be narrow band and it may limit potential applications.

GRIN phononic crystals and metamaterials can be applied to various types of waves in a long frequency limit, such as surface water waves[35, 36] for a few Hz, air-borne sound waves[19, 37-46] for $10^3$Hz-$10^5$Hz, water-borne acoustic waves[20, 27, 47-49] for $10^4$Hz-$10^6$Hz, Rayleigh waves[50-54] for 10Hz-$10^8$Hz, Lamb waves[23-25, 28, 55-67] for $10^3$-$10^8$Hz, among others, with functionalities like focusing[18, 19], waveguiding[65, 68], mirage[69], beam splitting or deflection[36, 57, 58], cloaking [50, 59] or energy harvesting[66, 67, 70]. It is in the domain of elastic waves in plates where the most interesting applications at the nanoscale are found, since from the technological point of view low-dimensional materials are more interesting than bulk materials. The excellent compatibility of phononic crystals and metamateials with the nanoelectromechanical systems (NEMS) has been proven for applications in wireless telecommunications, sensing or thermal control, among others[71-76]. However, the propagation of elastic waves in plates is in general composed of three polarizations, which travel at different speeds and for which refractive devices designed for one of these polarizations will not work for the other two. Recently, the designs of GRIN devices based on phononic crystal plates have demonstrated that it is possible the simultaenous control of all fundamental modes propagating in thin elastic plates in a broadband frequency region[25, 26, 58].

The objective of this review is to provide a comprehensive picture of the evolution of the domain of GRIN devices for mechanical waves, to demonstrate the wide variety of applications at the nanoscale that these devices offer and to present the challenges to be accomplished to further develop this field. The paper is organized as follows: after this introduction, we will review the fundamental ideas of the homogenization of sonic and phononic crystals in Section II and of phononic crystal plates in Section III, which offers efficient tools to obtain the effective elastic properties and, consequently, their effective refractive index; Then we will review GRIN devices for bulk acoustic and elastic waves in Section IV and for flexural waves in Section V; the advanced full control of polarizations in elastic plates with multimodal GRIN devices will be reviewed in Section VI. Last section will present some conclusions and future challenges.

## 2. Homogenization of sonic and phononic crystals

Graded materials with specific variations of the refractive index are obviously not found in nature, and they have to be artificially engineered. The inclusion of scatterers in a given matrix is an excellent way for the realization of artificially graded materials, since the average behavior of the scatterers is to modify the effective velocity of acoustic waves in the matrix, and this effective velocity can be tuned by means of the size of the inclusions. When these inclusions are arranged in a regular array we call the composite a sonic (fluid matrix) or phononic (solid matrix) crystal, and the computation of the effective sound velocity is made by a set of mathematical tools called homogenization theories. Roughly speaking, the homogenization region is the range of frequencies at which the propagating field cannot distinguish the individual scatterers of the composite and perceives the structure as a uniform material with some effective parameters (stiffness constant, mass density, viscosity, etc. are the parameters to be computed by means of the most adequate homogenization method).

Homogenization is an old problem and a great bibliography is available in this realm (see for instance[77]). Our aim here is not to review all these methods, but to present some examples of these methods applied to the specific case of sonic and phononic crystals.

### 2.1 Homogenization of sonic crystals by multiple scattering theory

As mentioned before, a sonic crystal consists of a periodic distribution of solid or fluid inclusions embedded in a fluid matrix. When the operating wavelength is larger than the typical distance between inclusions (lattice constant) the field cannot distinguish individual scatterers and perceives the structure as a homogeneous material with some effective parameters (mass density, compressibility, speed of sound, etc). There exists a vast literature about homogenization of periodic and random materials[77], but it has been shown that multiple scattering theory is a versatile method in the case of sonic crystals. For instance, for the simple case of a two-dimensional arrangement of fluid inclusions (parameters labeled with "a") in a fluid background (parameters labeled "b"), the effective bulk modulus $B_{\text{eff}}$ and mass density $\rho_{\text{eff}}$ are given by

$$\frac{1}{B_{eff}} = \frac{f}{B_a} + \frac{1-f}{B_b}, \frac{\rho_{eff}}{\rho_b} = \frac{\rho_a(\Delta+f)+\rho_b(\Delta-f)}{\rho_a(\Delta-f)+\rho_b(\Delta+f)} \qquad \text{Eq.(1)}$$

where f is the filling fraction of the crystal (area of the inclusions divided by the area of the unit cell) and $\Delta$ represents a modification of the filling fraction due to the multiple scattering processes[78, 79].

For airborne propagation any solid material will have a much higher density than that of air, and the extreme impedance mismatch blocks the penetration of the acoustic wave into the solid inclusions. Therefore, those solids can be assumed to be "rigid" inclusions and their mass density and compressibility can be set as infinite. In this case, if the filling fraction is small and the multiple scattering interactions can be neglected( $\Delta$ can be regarded as 1) the effective mass density and velocity can be simplified as[78]

$$\rho_{eff} = \frac{\rho_{air}(1+f)}{1-f}, c_{eff} = c_{air}/\sqrt{1+f} \qquad \text{Eq.(2)}$$

and we recover the effective mass density and effective sound velocity derived by Barryman[80] and the heuristic model[81], respectively.

As an example of application, a circular cluster of 151 wooden cylinders arranged in a hexagonal lattice embedded in air was homogenized by multiple scattering, and the resulting effective parameters as a function of the filling fraction are shown in the left panel of Fig.1. We can clearly observe, especially from the insets, that the simplified approach by Eq. (2) for effective mass density is valid when the filling fraction is no larger than 0.7, meanwhile the maximum limit of filling fraction is 0.6 for effective sound velocity. Experiments verified the homogenization method, and it was shown that for wavelengths larger than four times the lattice constant, the cluster could be considered as a homogeneous cylinder with fluid-like properties[78].

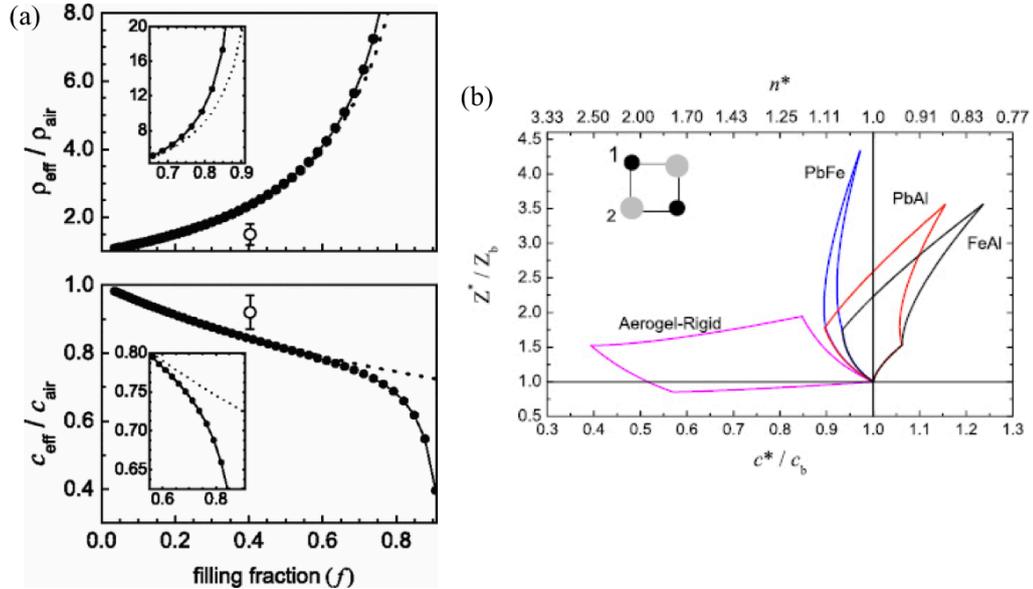

Figure 1. (a)Effective mass density and effective sound velocity for a circular cluster consists of 151 rigid cylinders embedded in air. The black dots stand for the results from the multiple scattering simulations Eq.(1) and the dashed lines are values obtained from Eq.(2)[78]; (b)Impedance-velocity diagram of sonic crystals consist of square lattice of solids 1 and 2 cylinders in air. Insert shows the unit cell[82].

The theory can be generalized to more cylinders per unit cell. If we consider two types of cylinders, with materials labeled as 1 and 2, embedded in a square lattice, their fluid-like parameters being ( $\rho_{a1}$, $B_{a1}$ ) and ( $\rho_{a2}$, $B_{a2}$) and the filling fractions being $f_1$ and $f_2$, respectively, the effective parameters can be obtained as

$$\frac{1}{B_{eff}} = \frac{1-f}{B_b} + \frac{f_1}{B_{a1}} + \frac{f_2}{B_{a2}}, \ \rho_{eff} = \frac{1+f_1\eta_1+f_2\eta_2}{1-f_1\eta_1-f_2\eta_2}\rho_b, c_{eff} = \sqrt{\frac{B_{eff}}{\rho_{eff}}} \qquad \text{Eq.(3)}$$

where $f = f_1 + f_2$, $\eta_i = (\rho_i - \rho_b)/(\rho_i + \rho_b)$.

The right panel of Fig.1 shows a "phase diagram" of the effective impedance and effective sound velocity for several selected pairs of materials under the low filling fraction condition (neglecting multiple scattering interactions), for different pair of materials and backgrounds (Pb-Fe, Pb-Al, Fe-Al in water and Aerogel-rigid cylinders in air). Any point on each curve is calculated sweeping over the total filling fraction. The overlapped corner point (1, 1) in this diagram corresponds to the case of $f_1$=0 and $f_2$=0. Notice that the effective impedance matches that of background when the sonic crystal is made of Aerogel and rigid cylinders embedded in air, making the sonic crystal transparent to air possessing a different refractive index. This interesting property was used in ref[17] to desin a completely transparent GRIN lens.

Multiple scattering allows the homogenization of finite clusters, therefore it is a versatile method to study the behavior of the effective parameters when disorder is introduced in the lattice[83] or even the effect of the cluster's size, as it was done in ref[84], where it was found that small clusters with specific number of scatterers presented the same effective parameters than that of the infinite medium.

## 2.2 Homogenization of periodic phononic crystals

Metamaterials for acoustic or elastic waves have been mainly studied by means of sonic or phononic crystals. As said before, in the low frequency limit, an anisotropic phononic crystal behaves as a homogeneous material with some effective parameters. The plane wave expansion (PWE) method is an alternative method to multiple scattering for the homogenization of periodic composites. This approach was firstly proposed by Krokhin[85] for sonic crystals and generalized for non-local phononic crystals in[86]. The equation of motion of an homogeneous elastic material assuming harmonic time-dependence is given by[87]

$$\rho\omega^2 u_i = k^2 n_{il} C_{IJ} n_{Jj} u_j \qquad \text{Eq.(4)}$$

where $\rho$ is the mass density, $\omega$ is the angular frequency, $u_i$ is the displacement, $k$ is the wavenumber, $\boldsymbol{n}$ is a unit vector, $\boldsymbol{C}$ is the stiffness tensor and we use Voigt notation for the sub indexes.

In a phononic crystal the mass density and stiffness tensors are periodic functions of the spatial coordinates, and an inhomogeneous version of the above equation has to be used. The plane wave expansion method is applied and then we arrive to an eigenvalue equation[88]

$$\omega^2 \rho_{\boldsymbol{G}-\boldsymbol{G'}}(u_{\boldsymbol{G'}})_i = (\boldsymbol{k}+\boldsymbol{G})_{iI} C_{IJ}^{\boldsymbol{G}-\boldsymbol{G'}}(\boldsymbol{k}+\boldsymbol{G'})_{Jj}(u_{\boldsymbol{G'}})_j \qquad \text{Eq.(5)}$$

where $\rho_{\boldsymbol{G}}$, $C_{IJ}^{\boldsymbol{G}}$ and $(u_{\boldsymbol{G'}})_i$ are the Fourier components of the mass density, stiffness tensor and the displacement field, respectively. The summation over repeated indexes has been assumed. The above equation can be manipulated to solve for the average field ($\boldsymbol{G}=0$ component), and this average field can be used to describe the effective behavior of the crystal. Then, after some complex mathematical manipulation we arrive to an expression for the dispersion relation of the average field, which consists in calculating the roots of the determinant of the matrix $\Gamma$ as

$$\Gamma_{ij} = \omega^2 \rho_{ij}^* - k^2 n_{iI} C_{IJ}^* n_{Jj} - \omega k (n_{iI} S_{Ij} + S_{ij}^\dagger n_{Jj}) \qquad \text{Eq.(6)}$$

where the effective mass density $\rho_{ij}^*$, the effective stiffness $C_{IJ}^*$ and the coupling tensor $S_{Ij}$ are[88]

$$\rho_{ij}^*(\omega,k) = \bar{\rho}\delta_{ij} + \omega^2 \rho_{-\boldsymbol{G'}} \chi_{ij}^{\boldsymbol{G'G}}(\omega,k)\rho_{\boldsymbol{G}} \qquad \text{Eq.(7a)}$$

$$C_{IJ}^*(\omega,k) = \bar{C}_{IJ} - C_{IL}^{-\boldsymbol{G'}}(k+\boldsymbol{G'})_{LI} \chi_{Im}^{\boldsymbol{G'G}}(\omega,k)(k+\boldsymbol{G})_{mM} C_{MJ}^{\boldsymbol{G}} \qquad \text{Eq.(7b)}$$

$$S_{Ij}(\omega,k) = \omega C_{IL}^{-\boldsymbol{G'}}(k+\boldsymbol{G'})_{LI} \chi_{Ij}^{\boldsymbol{G'G}}(\omega,k)\rho_{\boldsymbol{G}} \qquad \text{Eq.(7c)}$$

The above formulas offer a way to describe resonant ($\chi_{GG'}$) and nonlocal ($k$) phononic crystals. In the low frequency ($\omega \to 0$) and local ($k \to 0$) limit, from the Eq.(7), the coupling tensor $S_{Ij}$ =0; the mass density is a scalar as the volume average value, the effective stiffness tensor is simplified as

$$\rho_{ij}^* = \bar{\rho}\delta_{ij}, C_{IJ}^* = \bar{C}_{IJ} - C_{IL}^{-\boldsymbol{G'}}\boldsymbol{G'}_{LI}(M_{\boldsymbol{G'G}}^{-1})_{Im}^{\boldsymbol{G'G}}\boldsymbol{G}_{mM} C_{MJ}^{\boldsymbol{G}} \qquad \text{Eq.(8)}$$

where $(M_{\boldsymbol{G'G}})_{ij} = G_{iI} C_{IJ}^{\boldsymbol{G}-\boldsymbol{G'}} G'_{Jj}$. The information about the details of the phononic crystal structures is included in the stiffness tensor whose symmetry relates with the background matrix, inclusion, and lattice symmetry.

Periodic homogenization is therefore a versatile calculation method of the effective parameters, where once we know the shape of the scatterer or scatterers in the unit cell we can Fourier-transform their spatial distribution and obtain the effective parameters. The major drawback of this method is that it is not suitable for high-contrast of inclusions or fluid-elastic composites, since in this case convergence is poor. Also, due to periodicity, order-disorder effects are not easy to study. It is therefore a complementary technique to the multiple scattering method presented in the previous section.

## 3. Homogenization of phononic crystal plates

In a homogeneous thin elastic plate, the propagation of flexural waves (antisymmetric Lamb mode wave) can be approximately described by the bi-Helmholtz equation (assuming time harmonic dependence of the field)

$$(D_b \nabla^4 - \rho_b h_b \omega^2) W(x, y) = 0 \qquad \text{Eq.(9)}$$

where $\rho_b$, $h_b$ and $D_b = E_b h_b^3 / 12(1 - v_b^2)$ are the mass density, the thickness and the rigidity of the plate, respectively, being $E_b$ the Young's modulus and $v_b$ the Poisson's ratio. $W$ is the out of plane displacement field. Multiple scattering can also be applied for the homogenization of distribution of scatterers in elastic plates, and frequency-dependent effective parameters were obtained in[56]. The expressions derived there can also be applied in the low frequency limit to obtain the following effective parameters

$$\rho_{eff} = (1 - f)\rho_b + f\rho_a \qquad \text{Eq.(10a)}$$

$$D_{eff}(1 + v_{eff}) = \frac{(1+v_b)[D_b(1-v_b)+D_a(1+v_a)]-f(1-v_b)[D_b(1+v_b)-D_a(1+v_a)]}{D_b(1-v_b)+D_a(1+v_a)-f[D_b(1+v_b)-D_a(1+v_a)]} D_b \qquad \text{Eq.(10b)}$$

$$D_{eff}(1 - v_{eff}) = \frac{(1-v_b)[D_b(3+v_b)+D_a(1-v_a)]-f(3+v_b)[D_b(1-v_b)-D_a(1-v_a)]}{D_b(3+v_b)+D_a(1-v_a)-f[D_b(1-v_b)-D_a(1-v_a)]} D_b \qquad \text{Eq.(10c)}$$

where the subscript "a"("b") means the parameters for inclusions (background). The effective phase velocity can be obtained by the ratio of the wavenumber between the effective medium and the background, giving

$$\frac{c_{eff}}{c_b} = \frac{k_b}{k_{eff}} = \frac{(D_{eff}/\rho_{eff})^{1/4}}{(D_b/\rho_b)^{1/4}} \qquad \text{Eq.(11)}$$

As an example, let us consider circular inclusions in a thin aluminum plate ($h_b$=0.1$a$) and let the inclusion's material be chosen as hole (empty inclusion), lead and rubber, whose parameters can be found in Ref[56]. The effective parameters as given by Eqs.(10&11) as a function of the filling fraction are plotted in the left panel of Fig.2. The effective mass density is the volume average approach as for bulk waves. We see that the effective rigidity and the Poisson's ratio for the case of rubber behaves like that of hole, since the ratio of the Young's modulus between the rubber and the aluminum is very low. For the effective phase velocity, the difference between the rubber and the hole is more evident. In the right panel of Fig.2, it is shown the dispersion curves from finite element methods (red dots) and the effective parameters in Eq.(10)(blue lines) for a triangular lattice of hole/lead inclusions in a thin aluminum plate with the radius of inclusion being 0.3$a$. We can clearly observe that the agreement between the numerical simulation and the effective theory is quite good for the antisymmetric ($A_0$) and symmetric ($S_0$) Lamb modes. This is remarkable since we have developed the theory for the antisymmetric mode only, however there is a connection between the speed of these two modes[26] if a wave propagates from plate $h_1$ to plate $h_2$, $n_A^2 = n_S \frac{h_1}{h_2}$, which allow us to relate both, even if the propagation of the S mode has not been included in the theory.

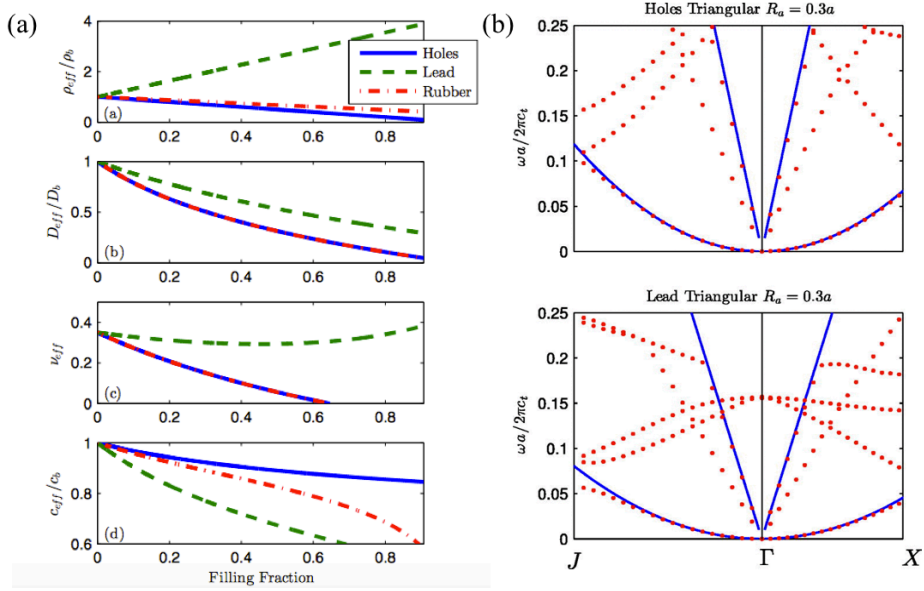

Figure 2. (a) The effective mass density, rigidity, Poisson's ratio and phase velocity with respect tot the filling fraction for different cylinder inclusions in an aluminum plate[56]; (b) Dispersion relations for a phononic crystal plate made of a triangular lattice of circular holes and lead inclusions with a radius of $0.3a$. the red dots show the band structures obtained by the finite element method while the blue lines show the dispersion curves calculated with homogenized parameters from Eq.(10)[89].

Actually, there are three fundamental plate modes as it is seen in the right panel of Fig.2, with the third one named shear-horizontal mode ($SH_0$), which needs an additional effective stiffness component $c_{66}^{eff}$ to compute its dispersion relation. The theory developed for flexural waves allows the calculation of the velocity of the S0 mode only due to the special relationship between the velocities of A0 and S0 modes, however a similar relationship is not found, or at least obvious, for the SH mode. However, the homogenization theory for bulk phononic crystals presented in Sec.2.2 offers the full components of the effective stiffness matrix, and it can be used to compute the effective velocity of shear waves in a phononic crystal plate. This connection is not obvious in principle, however it can be explained by a two-step homogenization procedure.

In Fig.3, upper-left panel, shows a two dimensional phononic crystal made of cylinders embedded in a bulk matrix. The phononic crystal plate at the upper-right panel can be regarded as a finite 'slice' taken from the bulk phononic crystal. In the low frequency limit, the phononic crystal behaves as a homogeneous material with effective parameters and tetragonal anisotropy (square lattice) or transversal isotropy (triangular lattice), and this symmetry is maintained for the phononic crystal plate, as illustrated from the upper panel to the lower panel in Fig.3(a). Thus, the effective phononic crystal plate can also be considered as a finite 'slice' of the effective homogeneous material with the same symmetry. Such homogenization procedure has a non-trivial implication in accordance with symmetry.

As an example, a phononic crystal plate with the unit cell shown in Fig.3(b) is considered for three different thicknesses, consisting of hole-gold shell unit cell in triangular lattice. The inner radius is $0.2a$ while the outer radius is $0.4a$. The dispersions obtained from homogenization method with effective parameters in Eq.(8) are compared to those from finite element methods, as displayed in Fig.3(b). In the low frequency limit, these effective phononic crystal plates behave like transversely isotropic plate. It is clear that the agreements between the finite element method and homogenization method are excellent for $S_0$ and $SH_0$ modes for all cases of thickness before the mode branches deviate, as their dispersions don't depend on the plate's thickness. However, the dispersion of $A_0$ mode relates with the plate's thickness. For thinner plates, the dispersion of fundamental $A_0$ mode keeps the parabolic shape throughout the $\Gamma$-J direction of the first irreducible Brillouin zone, so that there is an excellent agreement in the whole range of $\Gamma$-J for plate's thickness as $0.1a$. For thicker thickness such as $0.5a$ and $a$, the agreement for A0 mode goes well before the $A_0$ mode branch deviates.

As a conclusion, the homogenization procedure for phononic crystal plates illustrated in Fig.3(a) works well for low frequency limit and thin plates.

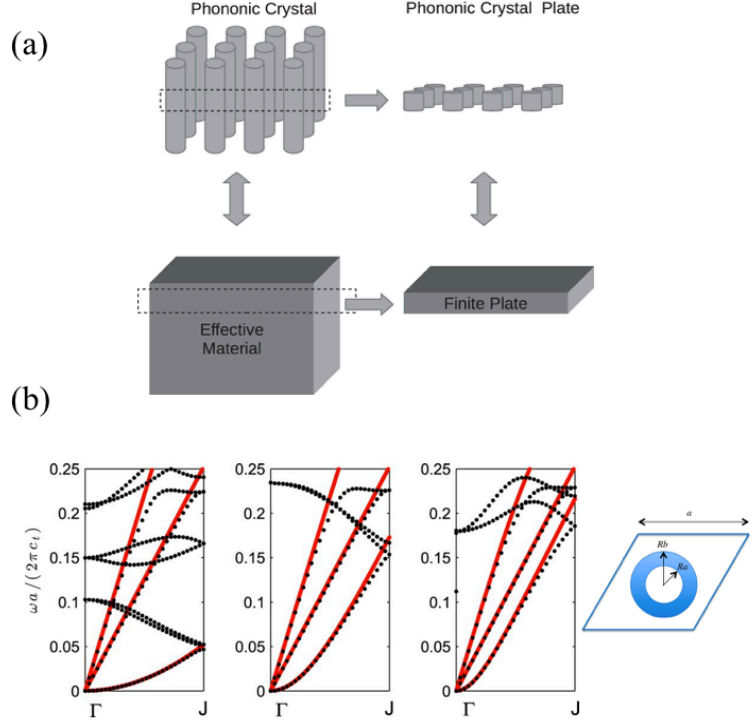

Figure 3. (a)Illustration of the homogenization relationship between phononic crystal and phononic crystal plate[26]; (b) Dispersion comparison of a phononic crystal plate between the finite element method (black dots) and the homogenization theory with effective parameters from Eq.(8) (red lines) for the plate thickness as $0.1a$(left), $0.5a$(middle) and $a$ (right). The phononic crystal plate consists of a triangular lattice of hole-gold shell unit cell whose inner radius is $0.2a$ and outer radius is $0.4a$[25].

## 4. GRIN devices for bulk acoustic and elastic waves

This section is devoted to the review of GRIN devices for waves in bulk materials, therefore no interfaces are assumed and surface or guided waves are excluded in this section (they will be considered in the next one). In principle we could divide the devices in two types, depending on the nature of the matrix which can be fluid or solid. This division makes sense since in the former shear waves does not exist in the background and the field is described entirely by a scalar pressure field, while in the latter shear waves with different propagation velocities have to be as well considered. However, for the fluid matrix an additional division can be made, since the physics involved in the scattering process is different if the background is air or water. While in air almost all solid materials are acoustically rigid due to the impedance mismatch between air and any solid, in underwater acoustics this impedance mismatch is of the same order, so that we have to consider the elastic nature of the inclusions. For this reason, we divide this section in three subsections, corresponding to the study and design of GRIN devices for air, water and solid backgrounds.

## 4.1. GRIN devices for bulk acoustic waves in air

In this section, we will discuss about the flat GRIN lens for focusing whose surface is flat instead of being curved as in traditional lens.

The index profile of GRIN flat lens keeps constant along the propagating $x$ axis while changes along the vertical $y$ axis as[18]

$$\text{n}(y) = n_0 \, \text{sech}(\alpha y) \, , \alpha = \frac{1}{h} cosh^{-1}\left(\frac{n_0}{n_h}\right) \qquad \text{Eq.(12)}$$

where h is the half-height in $y$ axis of the lens, $n_0$ is the refractive index at $y$=0, and $n_h$ is the index at the upper or lower edges $y = \pm$h. Such index profile is designed with the aim of presenting low aberrations[18]. The position of focal point $x_f$ at the right side of the lens can be derived from the differential equation satisfied by the energy rays[19].

As shown in Fig.4(a), a GRIN lens was fabricated with 9 columns of metal rods whose radii are varied locally by an inverse design as follows: first make a sweep of the filling fraction for sonic crystals made of rigid cylinders to plot the diagram of effective index and the filling fraction; then calculate the required index at a given position along y axis of the GRIN lens; next, choose the corresponding filling fraction of the rigid cylinder from the diagram in the first step in accordance with the required index in the second step; finally, calculate the required radius from the selected filling fraction in the third step. Figure 4(b) shows the pressure distribution fields after the GRIN sonic crystal at a frequency corresponding to a wavelength of about $3.8a$ numerically (upper panel) and experimentally (lower panel). Similar focal spots and side-lobes are evidently seen. To quantitatively characterize the focusing effect, the pressure along the $x$ and $y$ axes are also plotted in Fig.4(c), showing very good agreements. In this approach with rigid cylinders in GRIN sonic lens, the effective impedance of the lens is not matched with the background, so that the maximum pressure at the focal spot is limited. As indicated in Fig.1(b) in Sec.2.1, the mixed lattice of aerogel and rigid can also tailor the effective index while keeping the effective impedance matched with the background, so that zero reflectance can be achieved to make the lens transparent. The first GRIN sonic Wood lens proposed in Ref.[82] has a higher pressure focalization at the focal spot.

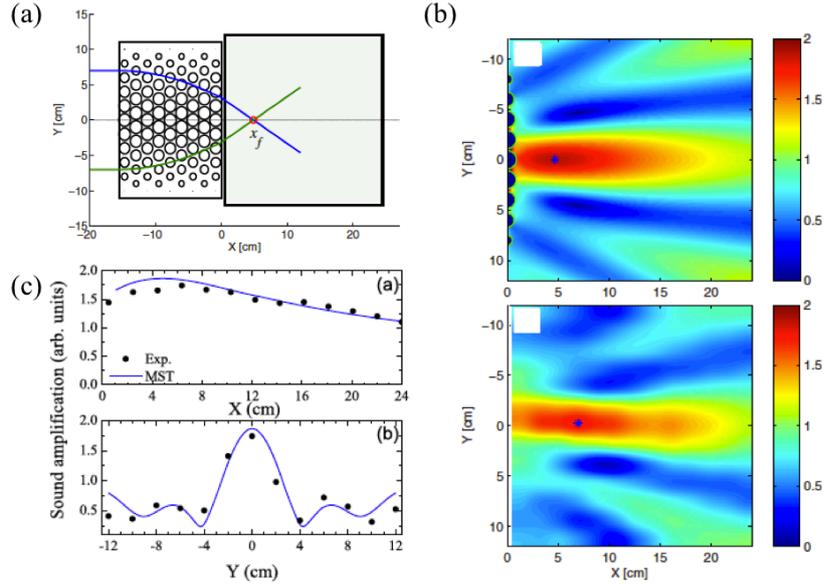

Figure 4. (a) Schematic view of a GRIN lens with the ray trajectory. For the impinging plane sound wave, the rays will bend and focus at $x_f$ position at the right side of the lens. (b) Normalized sound amplitude at 4.5kHz mapped at the right side of the GRIN lens from the multiple scattering method (upper panel) and the experiment (lower panel). (c) Normalized sound amplitude along the $x$ and $y$ axis crossing the focusing point. Blue lines stand for simulation and black dots stand for experiment.[19]

## 4.2. GRIN devices for bulk acoustic waves in water

Different from the impedance mismatch between the air and solid scatters, the mass density and acoustic velocity are in the same order for most scatterers and water, so that the scatters can't be regarded as rigid anymore. From Eq.(1) in Sec.2.1, one can also calculate the effective parameters of phononic crystals in water.

For the same type of GRIN flat lens as in the last section, similar phononic crystals made of radius-varied steel cylinders are proposed in water[90]. In the upper-left panel in Fig.5, the profile of radius is calculated to meet the require index profile like Eq.(12), being the maximum index at the center and minimum at the edges in vertical axis. A plane wave with center frequency at 20kHz is excited, corresponding to wavelength as 4 times lattice constant around the cutoff frequency in the low frequency limit. A clear focusing spot is observed from the experimental pressure intensity field. From the MST simulation, a similar position focal spot is also obtained as shown the dot '+'. To be noted that this GRIN lens also supports the source off the central axis, which was demonstrated numerically and experimentally[90].

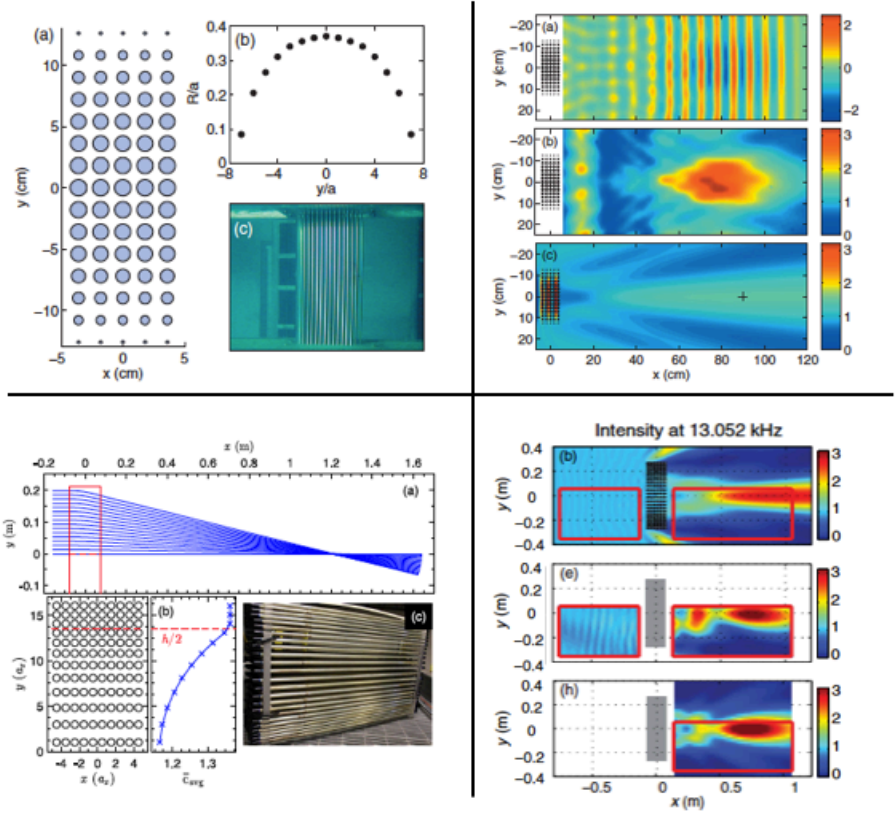

Figure 5. Upper-left panel: schematic view of GRIN flat lens with radius-varied steel cylinders embedded in water. Upper-right panel: normalized pressure amplitude from the source (a), measured normalized pressure intensity after the GRIN lens (b) and the same pressure intensity map using MST (c)[90]; Lower-left panel: schematic view of GRIN flat lens made of the same hollow air-filled aluminum tubes with anisotropic lattice spacing. Lower-right panel: Sound pressure intensity at 13.05kHz from 2D MST (top), experiment (middle) and 3D Rayleigh-Sommerfeld approximation (down)[20].

Eq.(12) gives a picture of special index profile for focusing with low aberration, while it does not define the maximum index at the center should be larger than that of the background, in another word, Eq.(12) can also be achieved by phase advance approach. In the lower panel of Fig.5, phononic crystals made of air-filled aluminum tubes are arranged in water with anisotropic lattice spacing. The thickness of the aluminum shell is 1/10 of the radius that is designed to have impedance matching with respect to water[91], a higher velocity but lower mass density acting as metafluids. The required filling fraction profile to meet the index distribution are implemented by using anisotropic lattice spacing[92]. From the lower-left panel of Fig.5, the lattice space in vertical axis is large at the center while reaches minimum at the upper and lower edges. The minimum velocity at the center is still larger than that of background water, resulting in a phase advance GRIN lens. A good comparison is found in the lower-right panel of Fig.5 among the experimental measurement and two predicted numerical simulations. It is worth mentioning that such GRIN flat lens is broadband, which allows to implement in water from experimental point of view as the short pulse from the transducer in water is normally broadband.

### 4.3. GRIN devices for bulk elastic waves

The effective parameters for bulk elastic wave in 2D phononic crystals can be found from Eq.(8) in Sec.2.2. Here, another approach based on the analysis of the lowest band among the dispersion curves is introduced. Let us consider a phononic crystal made of square lattice solid inclusions embedded in an elastic matrix. For small anisotropic ratios, the effective index for a bulk elastic wave can be obtained as[18]

$$n_{eff} = \frac{n_{\Gamma X} + n_{\Gamma M}}{2}, n_{\Gamma X} = \frac{c_b}{c_{\Gamma X}} = \frac{c_b}{d\omega/dk_{\Gamma X}}, n_{\Gamma M} = \frac{c_b}{c_{\Gamma M}} = \frac{c_b}{d\omega/dk_{\Gamma M}} \quad \text{Eq.(13)}$$

where $c_b$ is the acoustic velocity in the background matrix, $c$ is the group velocity of the bulk elastic wave mode, $\Gamma X$ and $\Gamma M$ are two orientations in the first Brillouin zone. For an epoxy matrix phononic crystal, the diagram of the first band of shear-vertical (SV) mode is plotted in two ways: i) steel inclusions with different filling fractions (different radii) as shown in the part (a) of the upper panel of Fig.6; ii) fixed filling fraction but with different material inclusions as shown in the part (b) of the upper panel of Fig.6. The elastic parameters for different solid materials can be found in Ref.[18]. Both approaches can change the slope of the first SV mode leading to a variation of effective index by Eq.(13).

From the dispersion curve variations, for a given band the effective index will change locally along the reduced frequency and for different bands the dispersive properties of the effective index are slightly different. Therefore, if a GRIN flat lens is designed based on a given frequency, it may have a small deviation in focal length as the effective or fitted gradient coefficient $\alpha$ in Eq.(12) changes, as seen in the lower panel of Fig.6.

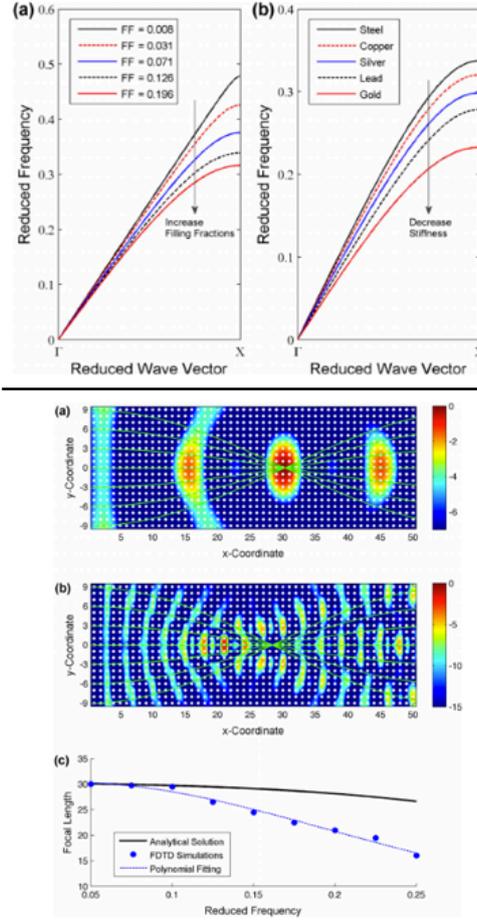

Figure 6. Upper panel:(a) The first shear-vertical (SV) mode of bulk elastic wave along the ΓX for varying the filling fractions of steel cylinders embedded in epoxy. The arrow indicates the increasing of the filling fraction. (b) Same band diagram but for phononic crystals made of different solid material inclusions in epoxy with a fixed filling fraction as 0.126. The arrow indicates the decreasing of stiffness. Lower panel: the FDTD simulated SV wave propagations in a phononic crystal whose inclusions are made of different materials embedded in epoxy at two different frequencies. The calculated and simulated focal lengths as a function of frequency are also plotted as solid line and dotted line, respectively[18].

## 5. GRIN devices for flexural waves

The control of elastic waves in plates provides the most interesting applications such as on-chip acoustic propagation, thermal conductivity control, opto-mechanic interactions in micro/nanoscale for GRIN devices, which can be attractive for the communities of physics, materials, and Nano science. For 2D thin plates, or membranes, the elastic waves propagate as surface waves due to the constrains of the plate's two boundaries, resulting in antisymmetric, symmetric and shear-horizontal polarizations. Among those modes, the fundamental antisymmetric mode, in another word flexural mode, is widely studied since it is mainly characterized by out-of-plane displacement component in plates that can be easily detected by means of optical methods.

Fig.7(a) shows the foci of the GRIN flat lens connected to a linear phononic crystal waveguide for flexural waves in a piezoelectric plate based on deep reactive ion etching. The well designed GRIN lens focus a plane incident wave as a spot at the interface between the GRIN lens and the phononic waveguide so that waveguiding is demonstrated. This would have potential for developing active micro plate lenses[93]. In Fig.7(b), GRIN flat lens is fabricated with silicon pillars erected on top to behave as a metalens to demonstrate the focusing behind the lens in the near-field beyond the diffraction limit. The dipolar resonant pillars and the flexural waves can exhibit polarization coherency to enhance the evanescent waves in order to help focal spot include better information[63]. The conception of mix local resonance and GRIN lens can also be applied to other types of waves. At nanoscale, a gradient optomechanical/phoxonic crystal in one dimensional semiconductor slab is fabricated with the minimum radius of the hole as only 100nm[73], as displayed in Fig.7(c). This phoxonic crystal demonstrate a full phononic bandgap at gigahertz domain. The optomechanical coupling originates from moving interfaces and the photoelastic effect. The acoustic mode inside the bandgap exhibits high mechanical quality factor making the optomechanical phonon coherent manipulation possible. Fig.7(d) illustrates the confinement of elastic waves in the GRIN devices can be used for energy harvesting[66, 67]. Piezoelectric energy harvester disks are deposited on a GRIN flat lens and a Luneburg lens at a position where the incident plane wave is focused as a spot. Comparing to the energy harvested in the background plate, the GRIN devices enable to generate output electrical power an order of magnitude higher.

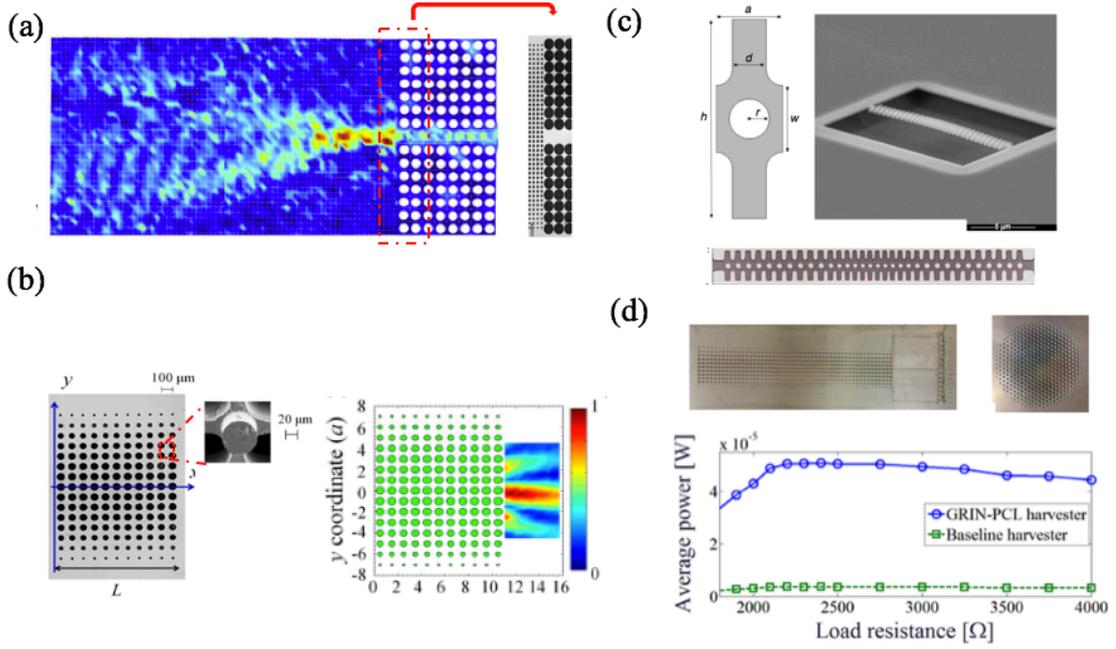

Figure 7. (a) Acoustic waveguide by combining GRIN flat lens and phononic crystals in a 80 μm thick plate. SEM image shows samples in the rectangular dotted box with the radii of gradient holes vary from 15μm to 25μm[93]; (b) acoustic metalens by combining GRIN flat lens and pillars (30 μm radius) to exhibit the near-field subdiffraction focusing[63]; (c) 1D phoxonic crystal with varied unit cells (the minimum radius of holes is 100 nm) to exhibit the optomechanical interactions[73]; (d) GRIN flat rectangular lens and Luneburg circular lens for energy harvesting[66, 67].

In Sec.3, Eq.(10) gives a useful tool to calculate the effective parameters of phononic crystal plates whose thickness is fixed as a constant. Therefore, GRIN devices can be designed by varying the filling ratio, in another word changing the radii of holes[57]. Now let us revisit the equation of motion (9) which describes flexural waves using Kirchoff-Love approximation. When a plane wave propagates in a thin plate with a wavenumber $k$, Eq.(9) gives the solutions for the dispersion relation and phase velocity as

$$k^4 = \frac{\rho h \omega^2}{D}, c^4 = (\frac{\omega}{k})^4 = \omega^2 \frac{E h^3}{12(1-v^2)\rho h} \qquad \text{Eq.(14)}$$

from where one can see that the phase velocity is not only related to the elastic parameters but also depends on plate's thickness. This special property of flexural waves allows to design GRIN devices just by locally varying the plate's thickness without any phononic crystal with holes. The refractive index for flexural wave can be easily derived when all elastic parameters remain fixed as in [55]

$$n_{eff} = \sqrt{\frac{h_b}{h_{eff}}}$$  Eq.(15)

In the upper-left panel of Fig.8, a set of omnidirectional GRIN devices are displayed with their refractive index profiles, namely Luneburg, Maxwell Fish-eye, 90 degree rotating, Eaton, Concentrator lenses[94, 95], as shown by the different colors. Implementing the refractive formula Eq.(15), their corresponding thickness variations are also shown in the panel below. The higher index requires, the thinner thickness behaves. For 90 degree rotating, Eaton and concentrator lenses, the maximum index at the center of the circular lens is larger than 5, so that the relative thickness there is smaller than 1/25. The designed GRIN lenses based on thickness variation are performed with full wave simulations and their corresponding wave functionalities are displayed at the bottom of Fig.8[24].

Invisibility is always a challenging problem for wave control in physics. Acoustic cloaks in principle require inhomogeneous and anisotropic materials for their realization, which are quite complicated to fabricate in practice[96-98]. A special gradient refractive index profile for omnidirectional invisible lens is derived that requires an infinite index in the vicinity of the center[99, 100]. However, such singularity can be achievable for flexural wave by Eq.(15). Nevertheless, from a practical point of view, the central index can be chosen with a proper finite value in accordance with the step in design. A ray enters the lens and exits in the same direction by flowing a loop around the center inside the lens, thus producing an invisible effect. The multi-frequency performance of the invisible lens is simulated with cylindrical waves from a point source[59]. To be noted that acoustic cloak for flexural wave based on nonlinear transformation acoustics method is proposed and realized also by varying the plate's thickness locally to fulfill the required rigidity profile[101].

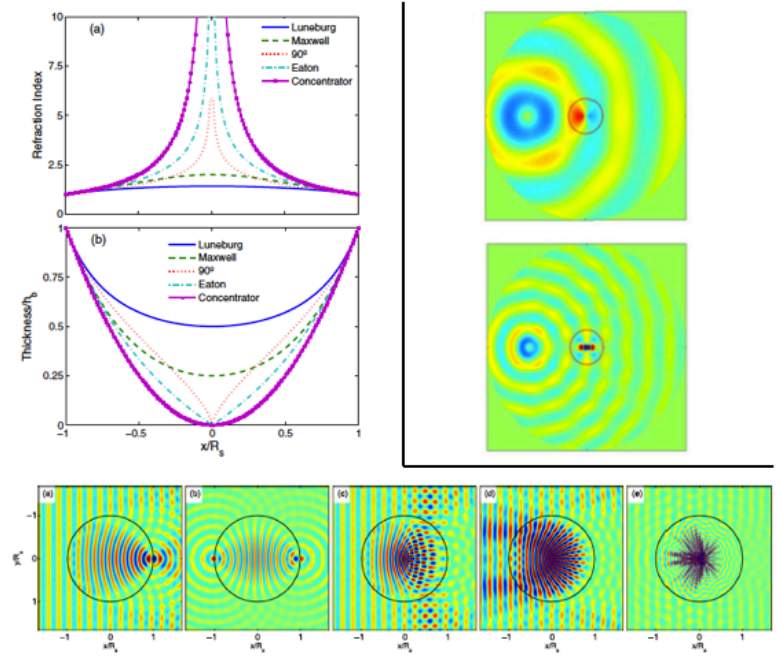

Figure 8. Upper-left panel: refractive index profiles and corresponding plate's thickness variations for Luneburg, Maxwell Fish-Eye, 90° rotating, Eaton, concentrator lenses. Wave performances of thoses lenses are displaced in the lower panel[24]. Upper-right panel: omnidirectional invisible lens for flexural waves based on thickness variation for two different frequencies[59].

## 6. Multimodal GRIN devices

From Fig.3(b), the fundamental Lamb modes have another two polarizations in addition to flexural mode. In fact, excitation in the thin plate from a given source will generate all kinds of modes, which means a design for only one mode may hinder the full functionalities of a GRIN device for applications such as absorption and energy harvesting. As discussed in the last section, the flexural wave can be independently controlled by thickness variation for given elastic parameters. The S0 and SH0 modes mainly depend on the elastic parameters ($c_{11}$, $c_{13}$, $c_{33}$) and ($c_{66}$), respectively, for transversely isotropic materials. Therefore, it needs a homogenization theory to obtain the full effective stiffness matrix. As illustrated in Fig.3 in Sec.3, the full homogenization method described in Sec.2.2 for bulk phononic crystal can be also applied to the thin phononic crystal plate by considering the latter as a finite thin 'slice' taken from the former. All the three polarizations of the Lamb waves are well described in the low frequency limit. Taking the same gold-shell-hole unit cell, we make a sweep for the outer radius meanwhile also varying the inner radius for a given outer radius. Figure 9(a) shows a full phase diagram of all possible pairs of ($n_{S0}$, $n_{SH0}$) domain with this sweeping approach, which exhibit several interesting properties: i) the dotted red line cutting the blue domain indicates that the effective indexes for the S0 and SH0 modes are identical, supporting the simultaneous control of these two modes; ii) the maximum effective index for S0 mode can reach up to almost 5 while it can be larger than 10 for SH0 mode; iii) the wide blue domain shows that it is possible to design a special ($n_{S0}$, $n_{SH0}$) trajectory that makes a GRIN device work as one type of lens for S0 mode and as another lens for SH0 mode. After defining the index profiles for SH0 and S0 modes, the index of A0 mode can be designed by means of an independent thickness variation, and such that it works similarly to the one or both of the SH0 and S0 modes or as another totally different lens. Thus, a full control of Lamb modes and advanced multimodal GRIN device become achievable.

In Fig.9(b), the red dotted circle shows the areas of Maxwell Fish-eye lens that is designed to work similarly for all the three fundamental modes. A point source is excited at the left border of the lens and a focused point image is formed at the right diametric border for a given frequency. It should be mentioned that the wavelengths of the three modes are different since their speeds are not the same due to the different dispersion property. The design of the full control means the positions of point focusing are the same, but not the wave field patterns. Finally, an advanced multimodal GRIN device is displayed in Fig.9(c): it works as Luneburg lens for the S0 and SH0 modes meanwhile it behaves as Maxwell Fish-eye lens for the A0 mode. The full control method is also extended to GRIN flat lens and beam splitter[58].

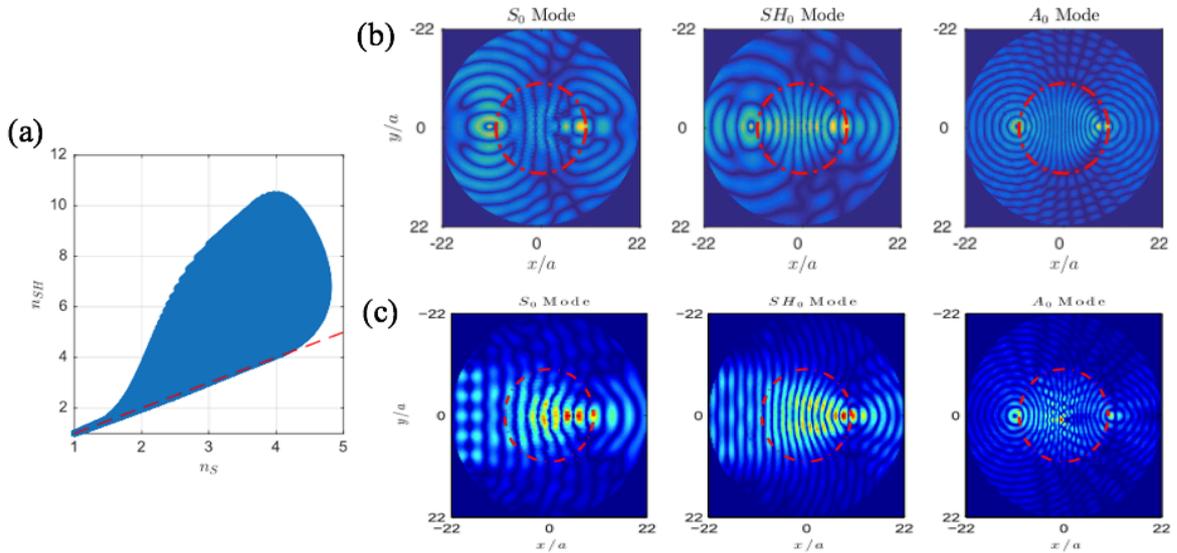

Figure 9. (a) Diagram of all possible values for the effective indexes for ($n_{S0}$, $n_{SH0}$) by varying the inner and outer radii of the gold-shell-hole structure. (b) GRIN Maxwell Fish-eye lens designed for all three Lamb modes simultaneously. (c) An advanced multimodal GRIN device works as Luneburg lens for the S0 and SH0 modes while as Maxwell Fish-eye lens for the flexural mode.[25]

## 7. Summary

Over the past two decades, we witnessed the rapid growth of the development of phononic crystals and metamaterials as well we the recent emergence of metasurfaces to deepen the understanding of wave physics. Although the effective index is a homogenized result from infinite periodic structures, it is still proper to define a local index in a gradient structure, being a bridge between infinity and locality. The homogenization theories of phononic crystals/metamaterials and the design principles of GRIN devices are implemented in terms of wavelength, mainly in long wavelength limit $\lambda > 3 \sim 4a$. Therefore, the gradient index conception can be applied to structure's size from nm to m and different wave natures propagating in solids, liquids and gases with free or constrained boundary conditions, among which the applications for elastic waves in thin plates at nanoscales show a promising potential in future for wireless telecommunications, heat conductivity, phonon sensor and so on. The complex properties of elastic waves in thin plates are also fully demonstrated to design more advanced and multimodal GRIN devices. Most applications of GRIN phononic crystals and metamaterials are dealing with lensing properties, while another big application is acoustic cloaking, which are studied among wide communities. Nevertheless, several challenges remain to be answered in future: 1) When wave propagates from one medium to another one, from a uniform surface point of view its behavior is governed by the Snell's law that involves the ratio of the refractive index between the two media. However, the wave energy transfer rate is governed by the impedance ratio between the two media. Some GRIN devices such as flat lens and transmitted-type metasurface, the impedance matching condition is not always easily fulfilled, which will limit the focusing energy level and the corresponding energy harvesting efficiency. 2) New refractive index profiles are needed to discover from mathematical methods that can help to design updated wave functionalities. 3) New physics such as thermal conductivity behaviors and nonlinear optical/acoustic responses in gradient phoxonic crystals at nanoscale. It is foreseeable that new breakthroughs will be reported by the united efforts in the communities of physics, material science, engineering, Nano science, mathematics among others.

**Acknowledgments**

Y.J. acknowledges a start-up fund from Tongji University. D.T. acknowledges financial support through the "Ramón y Cajal" fellowship and the U.S. Office of Naval Research under Grant No. N00014-17-1-2445.

**Reference**

[1] M.S. Kushwaha, P. Halevi, L. Dobrzynski, B. Djafari-Rouhani, Acoustic band structure of periodic elastic composites, Physical Review Letters, 71 (1993) 2022-2025.

[2] M. Sigalas, E.N. Economou, Band structure of elastic waves in two dimensional systems, Solid State Communications, 86 (1993) 141-143.

[3] Z.Y. Liu, X.X. Zhang, Y.W. Mao, Y.Y. Zhu, Z.Y. Yang, C.T. Chan, P. Sheng, Locally resonant sonic materials, Science, 289 (2000) 1734-1736.

[4] M.-H. Lu, L. Feng, Y.-F. Chen, Phononic crystals and acoustic metamaterials, Materials Today, 12 (2009) 34-42.

[5] Y. Pennec, J.O. Vasseur, B. Djafari-Rouhani, L. Dobrzyński, P.A. Deymier, Two-dimensional phononic crystals: Examples and applications, Surface Science Reports, 65 (2010) 229-291.

[6] M.I. Hussein, M.J. Leamy, M. Ruzzene, Dynamics of phononic materials and structures: Historical origins, recent progress, and future outlook, Applied Mechanics Reviews, 66 (2014) 040802.

[7] H. Ge, M. Yang, C. Ma, M.-H. Lu, Y.-F. Chen, N. Fang, P. Sheng, Breaking the barriers: advances in acoustic functional materials, National Science Review, 5 (2017) 159-182.

[8] P.A. Deymier, Acoustic metamaterials and phononic crystals, Springer Science & Business Media2013.

[9] A. Khelif, A. Adibi, Phononic Crystals: Fundamentals and Applications, Springer2015.

[10] L. Dobrzynski, A. Akjouj, Y. Pennec, H. Al-Wahsh, G. Lévêque, B. Djafari-Rouhani, Phononics: Interface Transmission Tutorial Book Series, Elsevier2017.

[11] R.V. Craster, S. Guenneau, Acoustic metamaterials: Negative refraction, imaging, lensing and cloaking, Springer Science & Business Media2012.

[12] S.A. Cummer, J. Christensen, A. Alù, Controlling sound with acoustic metamaterials, Nature Reviews Materials, 1 (2016) 16001.

[13] G. Ma, P. Sheng, Acoustic metamaterials: From local resonances to broad horizons, Science advances, 2 (2016) e1501595.

[14] Y. Jin, N. Fernez, Y. Pennec, B. Bonello, R.P. Moiseyenko, S. Hémon, Y. Pan, B. Djafari-Rouhani, Tunable waveguide and cavity in a phononic crystal plate by controlling whispering-gallery modes in hollow pillars, Physical Review B, 93 (2016) 054109.

[15] Y. Jin, D. Torrent, B. Djafari-Rouhani, Robustness of conventional and topologically protected edge states in phononic crystal plates, Physical Review B, 98 (2018) 054307.

[16] Y. Jin, B. Bonello, R.P. Moiseyenko, Y. Pennec, O. Boyko, B. Djafari-Rouhani, Pillar-type

acoustic metasurface, Physical Review B, 96 (2017) 104311.

[17] D. Torrent, J. Sánchez-Dehesa, Acoustic metamaterials for new two-dimensional sonic devices, New journal of physics, 9 (2007) 323.

[18] S.-C.S. Lin, T.J. Huang, J.-H. Sun, T.-T. Wu, Gradient-index phononic crystals, Physical Review B, 79 (2009) 094302.

[19] A. Climente, D. Torrent, J. Sánchez-Dehesa, Sound focusing by gradient index sonic lenses, Applied Physics Letters, 97 (2010) 104103.

[20] T.P. Martin, C.J. Naify, E.A. Skerritt, C.N. Layman, M. Nicholas, D.C. Calvo, G.J. Orris, D. Torrent, J. Sánchez-Dehesa, Transparent Gradient-Index Lens for Underwater Sound Based on Phase Advance, Physical Review Applied, 4 (2015) 034003.

[21] S. Peng, Z. He, H. Jia, A. Zhang, C. Qiu, M. Ke, Z. Liu, Acoustic far-field focusing effect for two-dimensional graded negative refractive-index sonic crystals, Applied Physics Letters, 96 (2010) 263502.

[22] Y. Tian, Z. Tan, X. Han, W. Li, Phononic crystal lens with an asymmetric scatterer, Journal of Physics D: Applied Physics, 52 (2019) 025102.

[23] X. Yan, R. Zhu, G. Huang, F.-G. Yuan, Focusing guided waves using surface bonded elastic metamaterials, Applied Physics Letters, 103 (2013) 121901.

[24] A. Climente, D. Torrent, J. Sánchez-Dehesa, Gradient index lenses for flexural waves based on thickness variations, Applied Physics Letters, 105 (2014) 064101.

[25] Y. Jin, D. Torrent, Y. Pennec, Y. Pan, B. Djafari-Rouhani, Gradient Index Devices for the Full Control of Elastic Waves in Plates, Scientific Reports, 6 (2016) 24437.

[26] Y. Jin, D. Torrent, Y. Pennec, Y. Pan, B. Djafari-Rouhani, Simultaneous control of the S 0 and A 0 Lamb modes by graded phononic crystal plates, Journal of Applied Physics, 117 (2015) 244904.

[27] A.S. Titovich, A.N. Norris, M.R. Haberman, A high transmission broadband gradient index lens using elastic shell acoustic metamaterial elements, The Journal of the Acoustical Society of America, 139 (2016) 3357-3364.

[28] K. Yi, M. Collet, M. Ichchou, L. Li, Flexural waves focusing through shunted piezoelectric patches, Smart Materials and Structures, 25 (2016) 075007.

[29] G. Yong, S. Hong-xiang, L. Chen, Q. Jiao, Y. Shou-qi, X. Jian-ping, G. Yi-jun, Z. Shu-yi, Acoustic focusing by an array of heat sources in air, Applied Physics Express, 9 (2016) 066701.

[30] B. Assouar, B. Liang, Y. Wu, Y. Li, J.-C. Cheng, Y. Jing, Acoustic metasurfaces, Nature Reviews Materials, DOI 10.1038/s41578-018-0061-4(2018).

[31] Y. Xu, Y. Fu, H. Chen, Planar gradient metamaterials, Nature Reviews Materials, 1 (2016) 16067.

[32] A. Kovalenko, M. Fauquignon, T. Brunet, O. Mondain-Monval, Tuning the sound speed in macroporous polymers with a hard or soft matrix, Soft Matter, 13 (2017) 4526-4532.

[33] A. Ba, A. Kovalenko, C. Aristégui, O. Mondain-Monval, T. Brunet, Soft porous silicone rubbers with ultra-low sound speeds in acoustic metamaterials, Scientific Reports, 7 (2017) 40106.

[34] Y. Jin, R. Kumar, O. Poncelet, O. Mondain-Monval, T. Brunet, Soft Metasurface with Gradient Acoustic Index, 11th International Conference of Electrical, Transport, and Optical Properties on Inhomogeneous Media, 2018, pp. 39.

[35] Z. Wang, P. Zhang, X. Nie, Y. Zhang, Focusing of liquid surface waves by gradient index lens, EPL (Europhysics Letters), 108 (2014) 24003.

[36] Z. Wang, P. Zhang, X. Nie, Y. Zhang, Manipulating water wave propagation via gradient index media, Scientific reports, 5 (2015) 16846.

[37] A. Climente, D. Torrent, J. Sánchez-Dehesa, Omnidirectional broadband acoustic absorber based on metamaterials, Applied Physics Letters, 100 (2012) 144103.

[38] Y.-J. Liang, L.-W. Chen, C.-C. Wang, I.-L. Chang, An acoustic absorber implemented by graded index phononic crystals, Journal of Applied Physics, 115 (2014) 244513.

[39] L.-Y. Wu, L.-W. Chen, An acoustic bending waveguide designed by graded sonic crystals, Journal of Applied Physics, 110 (2011) 114507.

[40] Y. Bao-guo, T. Ye, C. Ying, L. Xiao-jun, An acoustic Maxwell's fish-eye lens based on gradient-index metamaterials, Chinese Physics B, 25 (2016) 104301.

[41] R. Zhu, C. Ma, B. Zheng, M.Y. Musa, L. Jing, Y. Yang, H. Wang, S. Dehdashti, N.X. Fang, H. Chen, Bifunctional acoustic metamaterial lens designed with coordinate transformation, Applied Physics Letters, 110 (2017) 113503.

[42] Z. Zhang, R.-Q. Li, B. Liang, X.-Y. Zou, J.-C. Cheng, Controlling an acoustic wave with a cylindrically-symmetric gradient-index system, Chinese Physics B, 24 (2015) 024301.

[43] L. Zigoneanu, B.-I. Popa, S.A. Cummer, Design and measurements of a broadband two-dimensional acoustic lens, Physical Review B, 84 (2011) 024305.

[44] L.-Y. Wu, L.-W. Chen, Enhancing transmission efficiency of bending waveguide based on graded sonic crystals using antireflection structures, Applied Physics A, 107 (2012) 743-748.

[45] Y. Li, G. Yu, B. Liang, X. Zou, G. Li, S. Cheng, J. Cheng, Three-dimensional Ultrathin Planar Lenses by Acoustic Metamaterials, Scientific Reports, 4 (2014) 6830.

[46] V. Romero-García, A. Cebrecos, R. Picó, V.J. Sánchez-Morcillo, L.M. Garcia-Raffi, J.V. Sánchez-Pérez, Wave focusing using symmetry matching in axisymmetric acoustic gradient index lenses, Applied Physics Letters, 103 (2013) 264106.

[47] X. Su, A.N. Norris, C.W. Cushing, M.R. Haberman, P.S. Wilson, Broadband focusing of underwater sound using a transparent pentamode lens, The Journal of the Acoustical Society of America, 141 (2017) 4408-4417.

[48] R.A. Jahdali, Y. Wu, High transmission acoustic focusing by impedance-matched acoustic metasurfaces, Applied Physics Letters, 108 (2016) 031902.

[49] S.-D. Zhao, Y.-S. Wang, C. Zhang, High-transmission acoustic self-focusing and directional cloaking in a graded perforated metal slab, Scientific Reports, 7 (2017) 4368.

[50] A. Colombi, S. Guenneau, P. Roux, R.V. Craster, Transformation seismology: composite soil lenses for steering surface elastic Rayleigh waves, Scientific reports, 6 (2016) 25320.

[51] A. Colombi, D. Colquitt, P. Roux, S. Guenneau, R.V. Craster, A seismic metamaterial: The resonant metawedge, Scientific reports, 6 (2016) 27717.

[52] A. Colombi, V. Ageeva, R.J. Smith, A. Clare, R. Patel, M. Clark, D. Colquitt, P. Roux, S. Guenneau, R.V. Craster, Enhanced sensing and conversion of ultrasonic Rayleigh waves by elastic metasurfaces, Scientific Reports, 7 (2017) 6750.

[53] S. Jia-Hong, Y. Yuan-Hai, Beam focusing of surface acoustic wave using gradient-index phononic crystals, 2016 IEEE International Ultrasonics Symposium (IUS), 2016, pp. 1-3.

[54] J. Zhao, B. Bonello, L. Becerra, O. Boyko, R. Marchal, Focusing of Rayleigh waves with gradient-index phononic crystals, Applied Physics Letters, 108 (2016) 221905.

[55] A. Climente, D. Torrent, J. Sánchez-Dehesa, Omnidirectional broadband insulating device for flexural waves in thin plates, Journal of Applied Physics, 114 (2013) 214903.

[56] D. Torrent, Y. Pennec, B. Djafari-Rouhani, Effective medium theory for elastic metamaterials in thin elastic plates, Physical Review B, 90 (2014) 104110.

[57] D. Torrent, Y. Pennec, B. Djafari-Rouhani, Omnidirectional refractive devices for flexural waves based on graded phononic crystals, Journal of Applied Physics, 116 (2014) 224902.

[58] Y. Jin, D. Torrent, Y. Pennec, G. Lévêque, Y. Pan, B. Djafari-Rouhani, Multimodal and omnidirectional beam splitters for Lamb modes in elastic plates, AIP Advances, 6 (2016) 121602.

[59] Y. Jin, D. Torrent, B. Djafari-Rouhani, Invisible omnidirectional lens for flexural waves in thin elastic plates, Journal of Physics D: Applied Physics, 50 (2017) 225301.

[60] J. Zhao, R. Marchal, B. Bonello, O. Boyko, Efficient focalization of antisymmetric Lamb waves in gradient-index phononic crystal plates, Applied Physics Letters, 101 (2012) 261905.

[61] A. Zareei, A. Darabi, M.J. Leamy, M.-R. Alam, Continuous profile flexural GRIN lens: Focusing and harvesting flexural waves, Applied Physics Letters, 112 (2018) 023901.

[62] J. Zhao, B. Bonello, O. Boyko, Beam paths of flexural Lamb waves at high frequency in the first

band within phononic crystal-based acoustic lenses, AIP Advances, 4 (2014) 124204.

[63] J. Zhao, B. Bonello, O. Boyko, Focusing of the lowest-order antisymmetric Lamb mode behind a gradient-index acoustic metalens with local resonators, Physical Review B, 93 (2016) 174306.

[64] J. Zhao, B. Bonello, R. Marchal, O. Boyko, Beam path and focusing of flexural Lamb waves within phononic crystal-based acoustic lenses, New Journal of Physics, 16 (2014) 063031.

[65] T.-T. Wu, Y.-T. Chen, J.-H. Sun, S.-C.S. Lin, T.J. Huang, Focusing of the lowest antisymmetric Lamb wave in a gradient-index phononic crystal plate, Applied Physics Letters, 98 (2011) 171911.

[66] S. Tol, F.L. Degertekin, A. Erturk, Gradient-index phononic crystal lens-based enhancement of elastic wave energy harvesting, Applied Physics Letters, 109 (2016) 063902.

[67] S. Tol, F.L. Degertekin, A. Erturk, Phononic crystal Luneburg lens for omnidirectional elastic wave focusing and energy harvesting, Applied Physics Letters, 111 (2017) 013503.

[68] S.-C.S. Lin, R.B. Tittmann, J.-H. Sun, T.-T. Wu, T.J. Huang, Acoustic beamwidth compressor using gradient-index phononic crystals, Journal of Physics D: Applied Physics, 42 (2009) 185502.

[69] S.-C.S. Lin, T.J. Huang, Acoustic mirage in two-dimensional gradient-index phononic crystals, Journal of Applied Physics, 106 (2009) 053529.

[70] A. Darabi, M.J. Leamy, Analysis and Experimental Validation of an Optimized Gradient-Index Phononic-Crystal Lens, Physical Review Applied, 10 (2018) 024045.

[71] J. Cha, C. Daraio, Electrical tuning of elastic wave propagation in nanomechanical lattices at MHz frequencies, Nature Nanotechnology, DOI 10.1038/s41565-018-0252-6(2018).

[72] D. Hatanaka, I. Mahboob, K. Onomitsu, H. Yamaguchi, Phonon waveguides for electromechanical circuits, Nature Nanotechnology, 9 (2014) 520.

[73] J. Gomis-Bresco, D. Navarro-Urrios, M. Oudich, S. El-Jallal, A. Griol, D. Puerto, E. Chavez, Y. Pennec, B. Djafari-Rouhani, F. Alzina, A. Martínez, C.M.S. Torres, A one-dimensional optomechanical crystal with a complete phononic band gap, Nature Communications, 5 (2014) 4452.

[74] D. Navarro-Urrios, N.E. Capuj, M.F. Colombano, P.D. García, M. Sledzinska, F. Alzina, A. Griol, A. Martínez, C.M. Sotomayor-Torres, Nonlinear dynamics and chaos in an optomechanical beam, Nature Communications, 8 (2017) 14965.

[75] M.R. Wagner, B. Graczykowski, J.S. Reparaz, A. El Sachat, M. Sledzinska, F. Alzina, C.M. Sotomayor Torres, Two-Dimensional Phononic Crystals: Disorder Matters, Nano Letters, 16 (2016) 5661-5668.

[76] J. Cha, K.W. Kim, C. Daraio, Experimental realization of on-chip topological nanoelectromechanical metamaterials, Nature, 564 (2018) 229.

[77] G.W. Milton, The Theory of Composites, Cambridge University Press, Cambridge, UK, 2002.

[78] D. Torrent, A. Håkansson, F. Cervera, J. Sánchez-Dehesa, Homogenization of Two-Dimensional Clusters of Rigid Rods in Air, Physical Review Letters, 96 (2006) 204302.

[79] D. Torrent, J. Sanchez-Dehesa, Effective parameters of clusters of cylinders embedded in a nonviscous fluid or gas, Physical Review B, 74 (2006).

[80] J.G. Berryman, Long-wavelength propagation in composite elastic media I. Spherical inclusions, The Journal of the Acoustical Society of America, 68 (1980) 1809-1819.

[81] F. Cervera, L. Sanchis, J.V. Sánchez-Pérez, R. Martínez-Sala, C. Rubio, F. Meseguer, C. López, D. Caballero, J. Sánchez-Dehesa, Refractive Acoustic Devices for Airborne Sound, Physical Review Letters, 88 (2001) 023902.

[82] D. Torrent, J. Sanchez-Dehesa, Acoustic metamaterials for new two-dimensional sonic devices, New Journal of Physics, 9 (2007).

[83] D. Torrent, J. Sánchez-Dehesa, Effective parameters of clusters of cylinders embedded in a nonviscous fluid or gas, Physical Review B, 74 (2006) 224305.

[84] D. Torrent, J. Sánchez-Dehesa, F. Cervera, Evidence of two-dimensional magic clusters in the scattering of sound, Physical Review B, 75 (2007) 241404.

[85] A.A. Krokhin, J. Arriaga, L.N. Gumen, Speed of Sound in Periodic Elastic Composites, Physical Review Letters, 91 (2003) 264302.

[86] A.N. Norris, A.L. Shuvalov, A.A. Kutsenko, Analytical formulation of three-dimensional dynamic homogenization for periodic elastic systems, Proceedings of the Royal Society A: Mathematical, Physical and Engineering Sciences, 468 (2012) 1629-1651.

[87] D. ROYER, D.P. Morgan, E. Dieulesaint, Elastic Waves in Solids I: Free and Guided Propagation, Springer Berlin Heidelberg1999.

[88] D. Torrent, Y. Pennec, B. Djafari-Rouhani, Resonant and nonlocal properties of phononic metasolids, Physical Review B, 92 (2015) 174110.

[89] Y. Jin, D. Torrent, Y. Pennec, Y. Pan, B. Djafari-Rouhani, Simultaneous control of the S0 and A0 Lamb modes by graded phononic crystal plates, Journal of Applied Physics, 117 (2015) 244904.

[90] T.P. Martin, M. Nicholas, G.J. Orris, L.-W. Cai, D. Torrent, J. Sánchez-Dehesa, Sonic gradient index lens for aqueous applications, Applied Physics Letters, 97 (2010) 113503.

[91] T.P. Martin, C.N. Layman, K.M. Moore, G.J. Orris, Elastic shells with high-contrast material properties as acoustic metamaterial components, Physical Review B, 85 (2012) 161103.

[92] S.-C.S. Lin, B.R. Tittmann, T.J. Huang, Design of acoustic beam aperture modifier using gradient-index phononic crystals, Journal of Applied Physics, 111 (2012) 123510.

[93] M.-J. Chiou, Y.-C. Lin, T. Ono, M. Esashi, S.-L. Yeh, T.-T. Wu, Focusing and waveguiding of

Lamb waves in micro-fabricated piezoelectric phononic plates, Ultrasonics, 54 (2014) 1984-1990.

[94] M. Šarbort, T. Tyc, Spherical media and geodesic lenses in geometrical optics, Journal of Optics, 14 (2012) 075705.

[95] E.E. Narimanov, A.V. Kildishev, Optical black hole: Broadband omnidirectional light absorber, Applied Physics Letters, 95 (2009) 041106.

[96] A.C. Steven, S. David, One path to acoustic cloaking, New Journal of Physics, 9 (2007) 45.

[97] D. Torrent, J. Sánchez-Dehesa, Acoustic cloaking in two dimensions: a feasible approach, New Journal of Physics, 10 (2008) 063015.

[98] H. Chen, C.T. Chan, Acoustic cloaking and transformation acoustics, Journal of Physics D: Applied Physics, 43 (2010) 113001.

[99] J.C. Miñano, Perfect imaging in a homogeneous three-dimensional region, Optics Express, 14 (2006) 9627-9635.

[100] T. Tyc, H. Chen, A. Danner, Y. Xu, Invisible lenses with positive isotropic refractive index, Physical Review A, 90 (2014) 053829.

[101] A. Darabi, A. Zareei, M.R. Alam, M.J. Leamy, Experimental Demonstration of an Ultrabroadband Nonlinear Cloak for Flexural Waves, Physical Review Letters, 121 (2018) 174301.